\documentclass[%
 reprint,
superscriptaddress,
 amsmath,amssymb,
 aps,
pra,
]{revtex4-2}

\usepackage{graphicx}
\usepackage{dcolumn}
\usepackage{xcolor}
\usepackage{bm}
\usepackage{orcidlink}
\usepackage{hyperref}

\begin{document}


\title{Radiative lifetime of the $A$~$^2 \Pi_{1/2}$ state in RaF with relevance to laser cooling}

\author{M.~Athanasakis-Kaklamanakis\orcidlink{0000-0003-0336-5980}}
 \email{m.athkak@cern.ch}
\affiliation{Experimental Physics Department, CERN, CH-1211 Geneva 23, Switzerland}
\affiliation{KU Leuven, Instituut voor Kern- en Stralingsfysica, B-3001 Leuven, Belgium}
\affiliation{Centre for Cold Matter, Imperial College London, SW7 2AZ London, United Kingdom}

\author{S.~G.~Wilkins}
\affiliation{Department of Physics, Massachusetts Institute of Technology, Cambridge, MA 02139, USA}
\affiliation{Laboratory for Nuclear Science, Massachusetts Institute of Technology, Cambridge, MA 02139, USA}

\author{{P.~Lass\`egues}}
\affiliation{KU Leuven, Instituut voor Kern- en Stralingsfysica, B-3001 Leuven, Belgium}

\author{L.~Lalanne}
\affiliation{Experimental Physics Department, CERN, CH-1211 Geneva 23, Switzerland}

\author{J.~R.~Reilly}
\affiliation{Department of Physics and Astronomy, The University of Manchester, Manchester M13 9PL, United Kingdom}

\author{{O.~Ahmad}}
\affiliation{KU Leuven, Instituut voor Kern- en Stralingsfysica, B-3001 Leuven, Belgium}

\author{M.~Au\orcidlink{0000-0002-8358-7235}}
\affiliation{Systems Department, CERN, CH-1211 Geneva 23, Switzerland}
\affiliation{Department of Chemistry, Johannes Gutenberg-Universit\"{a}t Mainz, 55099 Mainz, Germany}

\author{{S.~W.~Bai}}
\affiliation{School of Physics and State Key Laboratory of Nuclear Physics and Technology, Peking University, Beijing 100971, China}

\author{{J.~Berbalk}}
\affiliation{KU Leuven, Instituut voor Kern- en Stralingsfysica, B-3001 Leuven, Belgium}

\author{{C.~Bernerd}}
\affiliation{Systems Department, CERN, CH-1211 Geneva 23, Switzerland}

\author{A.~Borschevsky\orcidlink{0000-0002-6558-1921}}
\affiliation{Van Swinderen Institute of Particle Physics and Gravity, University of Groningen, Groningen 9712 CP, Netherlands}

\author{A.~A.~Breier\orcidlink{0000-0003-1086-9095}}
\affiliation{Institut f\"ur Optik und Atomare Physik, Technische Universit\"at Berlin, 10623 Berlin, Germany}
\affiliation{Laboratory for Astrophysics, Institute of Physics, University of Kassel, Kassel 34132, Germany}

\author{K.~Chrysalidis}
\affiliation{Systems Department, CERN, CH-1211 Geneva 23, Switzerland}

\author{T.~E.~Cocolios\orcidlink{0000-0002-0456-7878}}
\affiliation{KU Leuven, Instituut voor Kern- en Stralingsfysica, B-3001 Leuven, Belgium}

\author{R.~P.~de~Groote\orcidlink{0000-0003-4942-1220}}
\affiliation{KU Leuven, Instituut voor Kern- en Stralingsfysica, B-3001 Leuven, Belgium}

\author{C.~M.~Fajardo-Zambrano\orcidlink{0000-0002-6088-6726}}
\affiliation{KU Leuven, Instituut voor Kern- en Stralingsfysica, B-3001 Leuven, Belgium}

\author{K.~T.~Flanagan\orcidlink{0000-0003-0847-2662}}
\affiliation{Department of Physics and Astronomy, The University of Manchester, Manchester M13 9PL, United Kingdom}
\affiliation{Photon Science Institute, The University of Manchester, Manchester M13 9PY, United Kingdom}

\author{S.~Franchoo}
\affiliation{Laboratoire Ir\`{e}ne Joliot-Curie, Orsay F-91405, France}
\affiliation{University Paris-Saclay, Orsay F-91405, France}

\author{R.~F.~Garcia~Ruiz}
\affiliation{Department of Physics, Massachusetts Institute of Technology, Cambridge, MA 02139, USA}
\affiliation{Laboratory for Nuclear Science, Massachusetts Institute of Technology, Cambridge, MA 02139, USA}

\author{D.~Hanstorp\orcidlink{0000-0001-6490-6897}}
\affiliation{Department of Physics, University of Gothenburg, Gothenburg SE-41296, Sweden}

\author{R.~Heinke}
\affiliation{Systems Department, CERN, CH-1211 Geneva 23, Switzerland}

\author{{P.~Imgram}\orcidlink{0000-0002-3559-7092}}
\affiliation{KU Leuven, Instituut voor Kern- en Stralingsfysica, B-3001 Leuven, Belgium}

\author{\'{A}.~Koszor\'{u}s\orcidlink{0000-0001-7959-8786}}
\affiliation{Experimental Physics Department, CERN, CH-1211 Geneva 23, Switzerland}
\affiliation{KU Leuven, Instituut voor Kern- en Stralingsfysica, B-3001 Leuven, Belgium}

\author{A.~A.~Kyuberis\orcidlink{0000-0001-7544-3576}}
\affiliation{Van Swinderen Institute of Particle Physics and Gravity, University of Groningen, Groningen 9712 CP, Netherlands}

\author{{J.~Lim}\orcidlink{0000-0002-1803-4642}}
\affiliation{Centre for Cold Matter, Imperial College London, SW7 2AZ London, United Kingdom}

\author{{Y.~C.~Liu}}
\affiliation{School of Physics and State Key Laboratory of Nuclear Physics and Technology, Peking University, Beijing 100971, China}

\author{{K.~M.~Lynch}\orcidlink{0000-0001-8591-2700}}
\affiliation{Department of Physics and Astronomy, The University of Manchester, Manchester M13 9PL, United Kingdom}

\author{{A.~McGlone\orcidlink{0000-0003-4424-865X}}}
\affiliation{Department of Physics and Astronomy, The University of Manchester, Manchester M13 9PL, United Kingdom}

\author{{W.~C.~Mei}}
\affiliation{School of Physics and State Key Laboratory of Nuclear Physics and Technology, Peking University, Beijing 100971, China}

\author{G.~Neyens\orcidlink{0000-0001-8613-1455}}
\email{gerda.neyens@kuleuven.be}
\affiliation{KU Leuven, Instituut voor Kern- en Stralingsfysica, B-3001 Leuven, Belgium}

\author{{L.~Nies}\orcidlink{0000-0003-2448-3775}}
\affiliation{Experimental Physics Department, CERN, CH-1211 Geneva 23, Switzerland}

\author{A.~V.~Oleynichenko\orcidlink{0000-0002-8722-0705}}
\affiliation{Affiliated with an institute covered by a cooperation agreement with CERN.}

\author{{A.~Raggio}\orcidlink{0000-0002-5365-1494}}
\affiliation{Department of Physics, University of Jyv\"{a}skyl\"{a}, Jyv\"{a}skyl\"{a} FI-40014, Finland}

\author{S.~Rothe}
\affiliation{Systems Department, CERN, CH-1211 Geneva 23, Switzerland}

\author{L.~V.~Skripnikov\orcidlink{0000-0002-2062-684X}}
\affiliation{Affiliated with an institute covered by a cooperation agreement with CERN.}

\author{{E.~Smets}}
\affiliation{KU Leuven, Instituut voor Kern- en Stralingsfysica, B-3001 Leuven, Belgium}

\author{B.~van~den~Borne\orcidlink{0000-0003-3348-7276}}
\affiliation{KU Leuven, Instituut voor Kern- en Stralingsfysica, B-3001 Leuven, Belgium}

\author{{J.~Warbinek}}
\affiliation{GSI Helmholtzzentrum f\"ur Schwerionenforschung GmbH, 64291 Darmstadt, Germany}
\affiliation{Department of Chemistry - TRIGA Site, Johannes Gutenberg-Universit\"at Mainz, 55128 Mainz, Germany}

\author{J.~Wessolek\orcidlink{0000-0001-9804-5538}}
\affiliation{Department of Physics and Astronomy, The University of Manchester, Manchester M13 9PL, United Kingdom}
\affiliation{Systems Department, CERN, CH-1211 Geneva 23, Switzerland}

\author{X.~F.~Yang\orcidlink{0000-0002-1633-4000}}
\affiliation{School of Physics and State Key Laboratory of Nuclear Physics and Technology, Peking University, Beijing 100971, China}

\author{the ISOLDE Collaboration}

\date{\today}

\begin{abstract}
The radiative lifetime of the $A$~$^2 \Pi_{1/2}$ ($v=0$) state in radium monofluoride (RaF) is measured to be {35(1)~ns}. The lifetime of this state {\color{black}and the related decay rate $\Gamma = 2.86(8)\times10^{7}$ s$^{-1}$ are of relevance to the laser cooling of RaF via the optically closed $A$~$^2\Pi_{1/2} \leftarrow X$~$^2 \Sigma_{1/2}$ transition, which makes the molecule a promising probe to search for new physics.} RaF is found to have a comparable photon-scattering rate to {\color{black}homoelectronic} laser-coolable molecules. Thanks to its highly diagonal Franck-Condon matrix, it is expected to scatter an order of magnitude more photons {\color{black}than other molecules} when using just 3 cooling lasers, before it decays to a dark state. The lifetime measurement in RaF is benchmarked by measuring the lifetime of the $8P_{3/2}$ state in Fr to be 83(3)~ns, in agreement with literature.
\end{abstract}

\maketitle

\section{Introduction}
 The properties of the lowest-lying $^2 \Pi_{1/2}$ state in alkaline-earth monofluorides and other {\color{black}homoelectronic} molecules
 are of central importance for their suitability for direct laser cooling and trapping. The diagonal Franck-Condon matrix and closed optical cycle between the $X$~$^2 \Sigma^+$ ground state and the lowest-lying $A$~$^2 \Pi_{1/2}$ state of these molecules has enabled the direct laser cooling of molecules to mK temperatures or lower, such as CaF~\cite{Zhelyazkova2014}, SrF~\cite{Shuman2010}, YbF~\cite{Lim2018}, YO~\cite{Ding2020}, CaOH~\cite{Vilas2022}, SrOH~\cite{Kozyryev2017SrOH}, and the symmetric-top CaOCH$_3$~\cite{Mitra2020}.
 The {\color{black}short} radiative lifetime of the $A$~$^2 \Pi_{1/2}$ state is key for {\color{black}efficient} laser cooling based on the optically closed $A$~$^2 \Pi_{1/2} \leftarrow X$~$^2\Sigma^+$ transition,
{\color{black}and} decelerating the molecules to the capture velocity of a magneto-optical trap~\cite{Tarbutt2018}.

 {\color{black}The homoelectronic and radioactive RaF has been proposed as a sensitive probe for searches of parity or time-reversal ($P,T$) violating properties, such as the electric dipole moment of the electron (eEDM). Extending laser cooling and trapping to RaF would enable highly sensitive searches for new physics by significantly increasing the interrogation time of the RaF molecules.}

Thanks to the high atomic number of Ra ($Z=88$), the ground state of RaF, {\color{black}whose longest-lived isotopologue $^{226}$Ra$^{19}$F has a half-life of 1,600 years,} 
is highly sensitive to the eEDM ~\cite{Isaev2010,Wilkins2023BW}. 
Furthermore, the octupole-deformed shape of the $^{222-226}$Ra nuclei~\cite{Butler2020} also enhances the sensitivity of RaF to a number of {\color{black}$P,T$-violating} nuclear properties~\cite{Safronova2018}, such as the Schiff moment~\cite{Isaev2010,Kudashov2014}. {\color{black}Combined with the potential to achieve exceptionally long coherence times thanks to laser cooling, RaF is a promising system to search for new physics with potentially unprecedented sensitivity.} 


To determine the maximum photon scattering rate and the required interaction time for laser slowing of RaF, knowledge of the radiative lifetime of the upper state in the closed optical cycle is necessary. This work presents the experimental measurement of the radiative lifetime of the $A$~$^2 \Pi_{1/2}$ state in RaF at 35(1)~ns, via delayed multi-step ionization using broadband pulsed lasers ($>$3~GHz). Prior to this work, only an upper limit of $\tau<50$~ns was reported for this state~\cite{GarciaRuiz2020}.
{The result is benchmarked by applying the same measurement procedure also for the 8$P_{3/2}$ state of neutral Fr, for which a lifetime value was previously reported and lies in the same order of magnitude as that for $A$~$^2 \Pi_{1/2}$ in RaF. The benchmark measurement of 83(3)~ns in this study is consistent with the literature value of 83.5(15)~ns~\cite{Aubin2004} within 1$\sigma$.}

\section{Methods}
The measurements were obtained via delayed multi-step ionization with the Collinear Resonance Ionization Spectroscopy (CRIS) experiment using beams of $^{226}$Ra$^{19}$F$^+$ and $^{221}$Fr$^+$ produced at the ISOLDE radioactive ion beam facility at CERN. {\color{black}The technique and the employed laser schemes are shown schematically in Fig.~\ref{fig:FigRaF_2_1}, and further experimental details can be found in the Supplemental Material and in Refs.~\cite{Catherall2017,Cocolios2013CRIS,Au2023InsourceIntrap}.}

{\color{black}
Fast ions ($\sim$40~keV kinetic energy) of $^{226}$Ra$^{19}$F$^+$ and $^{221}$Fr$^+$, released as a pulsed beam from a linear Paul trap (100-Hz repetition, 5-$\mu$s temporal spread, 0.8-m longitudinal spatial spread), were delivered to the CRIS experiment and were neutralized via collisions with a vapor of sodium atoms in a charge-exchange cell, while the remaining ions were deflected onto a beam dump. The neutralized bunches entered the interaction region, where they were temporally and spatially overlapped in a collinear geometry with pulsed laser beams (see Fig.~\ref{fig:FigRaF_2_1}a) that step-wise excited the molecules/atoms from the ground state to above the ionization potential, as per the laser schemes shown in Fig.~\ref{fig:FigRaF_2_1}b,c. The resonantly re-ionized molecules were deflected onto a MagneToF single-ion detector, while the residual neutral beam was discarded onto a beam dump.}

\begin{figure}
    \centering
     \includegraphics[width=0.45\textwidth]{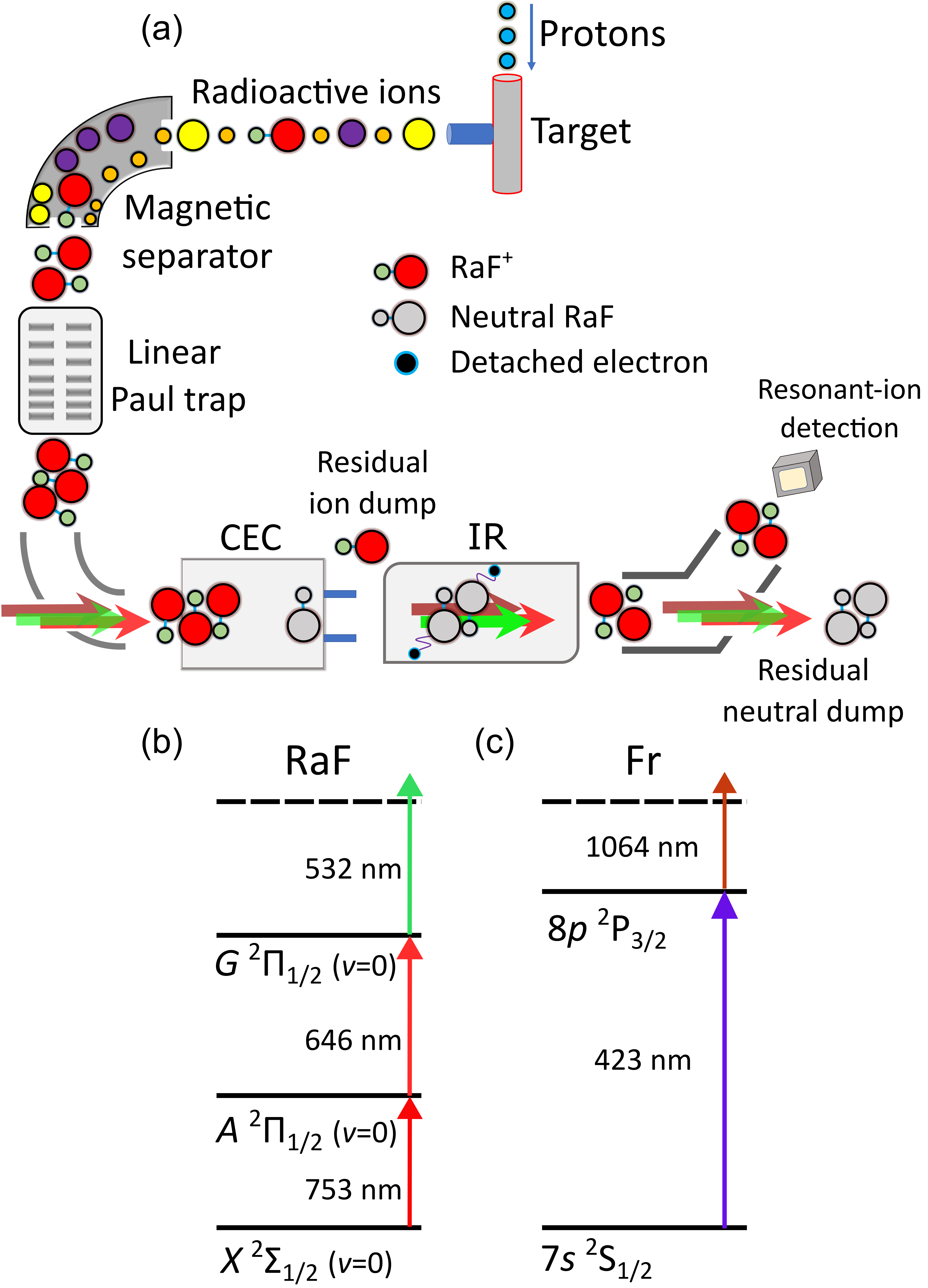}
    \caption{ {\color{black}Top: \textbf{(a)} Summary of the experimental setup, showing the production, mass separation, bunching in a Paul trap, neutralization in the charge-exchange cell (CEC), and resonant ionization in the interaction region (IR). Bottom: Laser excitation scheme used for \textbf{(b)} RaF and \textbf{(c)} Fr in this work, denoting the rest-frame transition wavelengths.}}
    \label{fig:FigRaF_2_1}
\end{figure}

All lasers in this study were pulsed and broadband, with linewidths $\Delta f \geq 3$~GHz. The 753-nm step in RaF was produced with the fundamental output of a titanium:sapphire (Ti:Sa) laser, while the 423-nm excitation step in Fr was produced by second-harmonic generation of the same Ti:Sa laser. The 646-nm step in RaF was produced using a dye laser pumped by a 532-nm Nd:YAG laser. The 532-nm and 1064-nm steps were produced by the second and fundamental harmonics of a Nd:YAG laser, respectively. The 646-nm and 532-nm steps in RaF and 1064-nm step in Fr are referred to as the ionization steps, while the 753-nm and 423-nm steps are referred to as the excitation steps in the respective schemes in this work.

The pulses of the laser used for the excitation steps had a full width at half-maximum (FWHM) of 38(2)~ns, while the pulses of the Nd:YAG lasers used for the ionization steps had a FWHM of 15(1)~ns. The {\color{black}shot-to-shot} jitter of the excitation step was measured to be 4~ns and that of the ionization steps to be 3~ns, for a total relative {\color{black}shot-to-shot} jitter of 5~ns {\color{black}(added in quadrature)} between excitation and ionization steps in both schemes {\color{black}(details in the Supplemental Material)}. As the 646-nm step in the RaF scheme was provided by a pulsed dye laser pumped by an identical Nd:YAG laser as the one used for the 532-nm ionization step, the FWHM and jitter of the second and third steps were measured to be the same, as expected.

%

To determine the lifetime of the $A$~$^2 \Pi_{1/2}$ state in RaF and the $8P_{3/2}$ state in Fr, multiple measurement cycles were performed for various timing arrangements between the ejection of the molecular/atomic beam from the gas-filled linear Paul trap of the ISOLDE facility~\cite{Mane2009} and the timing of the laser pulses.
Firstly, an optimal setting for the Paul trap ejection timing was determined based on the maximum laser-molecule/atom spatial and temporal overlap, and a lifetime measurement was taken for these settings, by moving the timing of the ionization steps forward in time while keeping the timing of the excitation step fixed. Lifetime measurements (denoted as Cycle 1 for both RaF and Fr) were performed for this optimal ejection timing, {\color{black}as well as} for ejection timings 0.5~$\mu$s earlier and 0.5~$\mu$s later compared to the optimal for Fr {\color{black}(1-$\mu$s steps for RaF)}. One lifetime measurement was also performed at the optimal ejection timing, but moving the excitation step timing backwards while keeping the timing of the ionization steps fixed. Only for the RaF experiment, an additional measurement (Cycle 4 in Fig.~\ref{fig:FigRaF_2_2}c) had been performed in a separate {\color{black}proof-of-principle} experimental campaign 2 years prior to the rest of the cycles; details about this measurement cycle can be found in the Supplemental Material.

{\color{black}Fig.~\ref{fig:FigRaF_2_2}a shows {\color{black}an example of} the measured ion count rate as a function of the delay time {\color{black}$t$} between the excitation step and {\color{black}the subsequent} ionization steps {\color{black}(all decay curves are shown in the Supplemental Material). For delay times larger than the sum of the pulse widths (FWHM) and the relative jitter, the data represents the radiative decay from the excited $A$~$^2 \Pi_{1/2}$ state, and the curve can then be fitted} with the natural decay law:

\begin{equation} \label{eq:decay_law}
    N(t) = N_0 e^{-t/\tau} + C
\end{equation}

In Eq.~\ref{eq:decay_law}, $N(t)$ is the population of molecules in the $A$~$^2 \Pi_{1/2}$ state as a function of the delay, $N_0$ is the population at delay time $t=0$, $\tau$ is the radiative lifetime of the $A$~$^2 \Pi_{1/2}$ state, and $C$ is a constant representing the ion background on the detector that stems from non-resonant ionization. {\color{black}In order to exclude data points where the radiative decay law is not valid (at times when there is an overlap of the lasers), the first 60 ns were omitted when fitting the decay curve, shown as the "laser overlap regime" in Fig.~\ref{fig:FigRaF_2_2}a. This span of 60 ns is approximately equal to the sum of the FWHM and relative jitter of the lasers (see Supplemental Material for details).}

{\color{black}The laser overlap regime was confirmed as shown in Fig.~\ref{fig:FigRaF_2_2}b, where the delay time at which the reduced $\chi^2$ saturates is verified, which marks the end of the laser overlap regime. While saturation already start from 40~ns, all decay curves were analyzed with a 60-ns start point to exclude data with partial laser overlap. Fitting the decay curves with start points between 40 to 70~ns resulted in consistent lifetime values. The same procedure was followed for Fr.}

{\color{black}A 5-ns error was adopted for the laser delay time, which corresponds to the combined shot-to-shot laser jitter of the excitation and ionization lasers (see Supplemental Material). Repeating the fit procedure with a smaller \textit{x}-error yielded a reduced $\chi^2$ progressively closer to 1, while the lifetime result remained unchanged. The uncertainties on the final lifetime values were not scaled with the respective reduced $\chi$ of each fit, as this would reduce the uncertainty, and it is preferred to remain conservative in the result.}

The fit residuals for the decay curve are shown in Fig.~\ref{fig:FigRaF_2_2}a. The fraction of data points whose deviation from the fit exceeds the $y$-error is consistent with the 1$\sigma$ confidence interval. {\color{black}A potential oscillatory trend in the residuals, not seen in all data sets (see Fig.~4 in Supplementary Material), falls within the confidence interval.}

\begin{figure}
    \centering
     \includegraphics[width=0.45\textwidth]{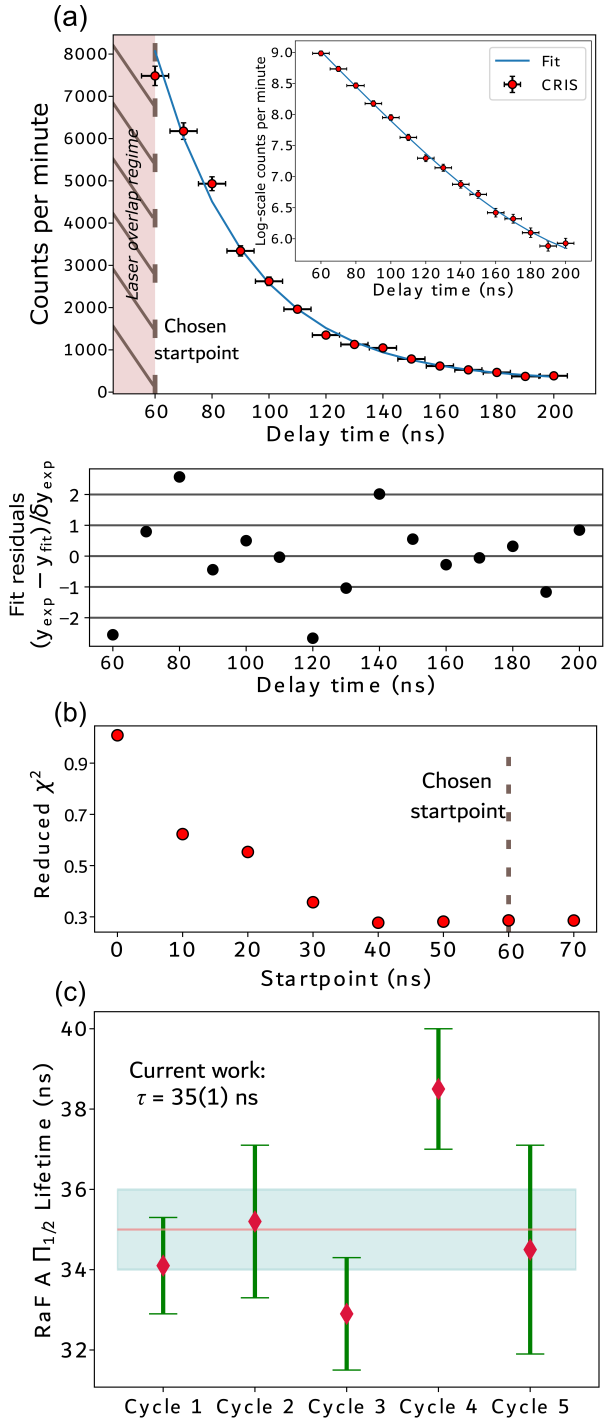}
    \caption{
    \textbf{(a)} Example decay curve {\color{black}(top) and fit residuals (bottom)} for a single measurement cycle in RaF. 
    The \textit{y}-errors include both statistical and systematic components, and the \textit{x}-error, the same for all points, corresponds to the total relative jitter of the lasers (see text and Supplemental Material for details). The inset shows the same plot with the \textit{y}-axis in logarithmic scale. 
    \textbf{(b)} Example curve of reduced-$\chi^2$ as a function of the fit start point in the data subset of the full decay curve for a single measurement cycle in RaF, used to confirm the start of free radiative decay (see text for details). \textbf{(c)} Extracted lifetime values for individual measurement cycles in RaF. The line and surrounding band denote the error-weighted mean and its standard deviation across all 5 measurement cycles, respectively.}
    \label{fig:FigRaF_2_2}
\end{figure}

{\color{black}In Fig.~\ref{fig:FigRaF_2_2}c, the final results from the lifetime analysis for RaF are presented, extracted from the different measurement cycles that cover different experimental conditions. The weighted mean value of the independent measurements is found to be 35(1) ns.}

To test the accuracy of the experimental approach and analysis procedure, the lifetime of the $8P_{3/2}$ state in Fr {\color{black}was measured using the same approach}, and compared with a literature value from Aubin \textit{et al.}~\cite{Aubin2004}. The results are presented in Fig.~\ref{fig:FigFrBenchmark}, {\color{black}leading to a weighted mean value} of 83(3)~ns, consistent with the literature value of 83.5(15)~ns~\cite{Aubin2004}. 


\section{Results and Discussion}
\begin{figure}
    \centering
    \includegraphics[width=0.45\textwidth]{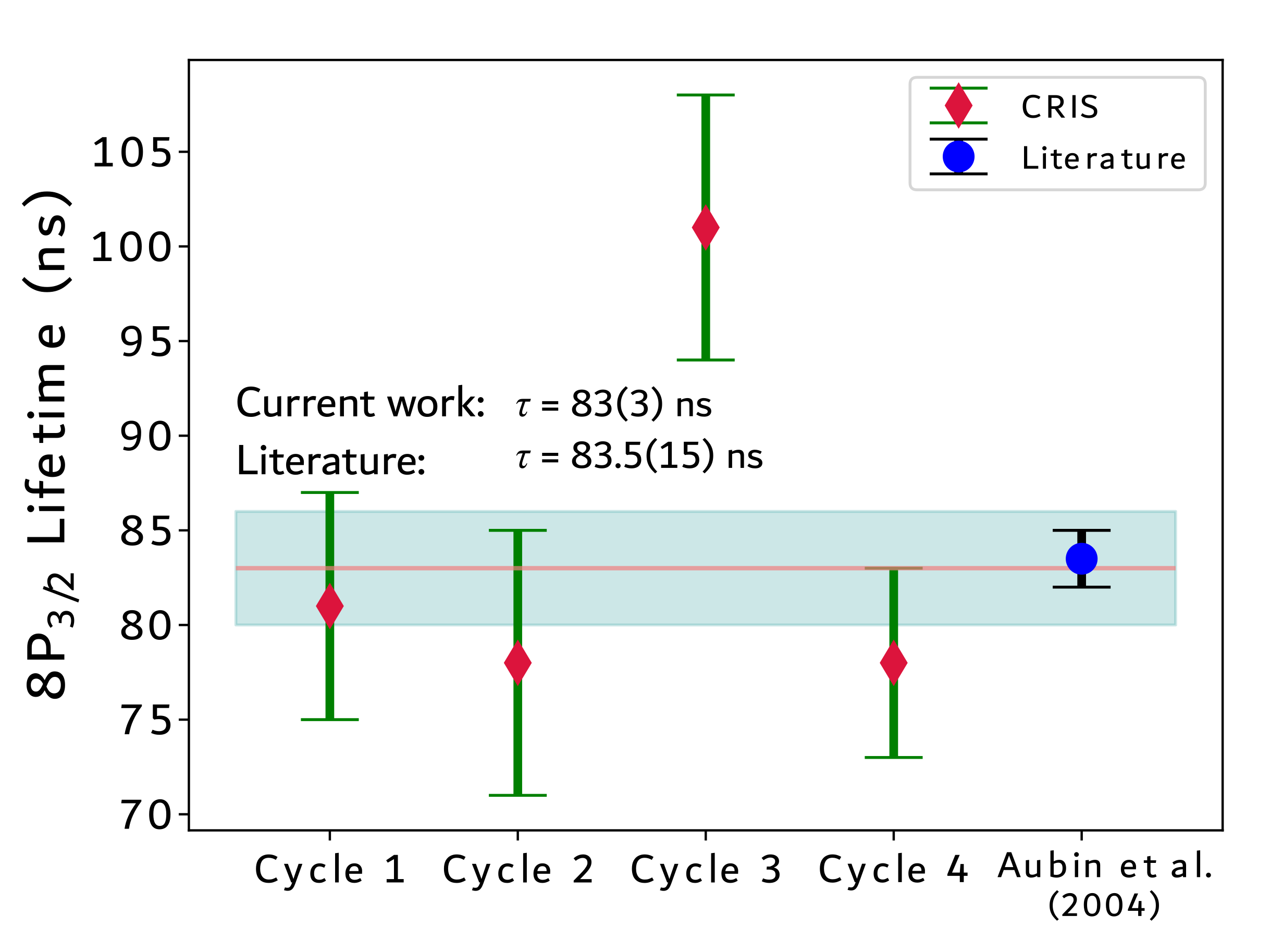}
    \caption{Lifetime results for the $8P_{3/2}$ state in Fr as a benchmark study. The central line and surrounding band represent the error-weighted mean and 1 standard deviation of the lifetime values of the individual measurement cycles. The final result of 83(3)~ns in this benchmark study is in agreement with the literature value from Ref.~\cite{Aubin2004}.}
    \label{fig:FigFrBenchmark}
\end{figure}

The lifetime of the $A$~$^2\Pi_{1/2}$ state in RaF was measured to be $\tau=35(1)$~ns. The radiative lifetime and excitation energy of the $A$~$^2 \Pi_{1/2}$ ($v=0$) states across alkaline-earth monofluorides (and the homoelectronic YbF) are shown in Fig.~\ref{fig:comparison}. A trend of gradually increasing lifetime for increasing molecular mass is evident, which appears correlated with a decrease in the excitation energy of the $A$~$^2 \Pi_{1/2}$ states. 

The excitation energy is of importance for the laser-cooling scheme, as {\color{black}it}
determines the single-photon recoil velocity, $\hbar k/ m$, where $m$ is the mass of the molecule, $k=\frac{2\pi}{\lambda}$, and $\lambda$ is the transition wavelength. The trends in Fig.~\ref{fig:comparison} imply that laser-cooling efficiency decreases for heavier alkaline-earth monofluorides, as the molecular mass and transition wavelength increase, while the increasing lifetime of the upper state leads to a reduced photon scattering rate. For the laser-cooling transition in RaF, the recoil velocity is equal to 0.22~cm~s$^{-1}$.

\begin{table}[]
\caption{Summary of present results on the radiative decay of the $A$~$^2 \Pi_{1/2}$ ($v=0$) state in RaF.}
\label{tab:summary}
\begin{tabular}{cc}
\hline\hline
Radiative lifetime $\tau$ (ns)  &  \hspace{4.5em}35(1)\\
Radiative decay rate $\Gamma$ (s$^{-1}$) &  \hspace{4.5em}2.86(8)$\times$10$^{7}$\\
Maximum scattering rate $R^{\rm{max}}_{\rm{sc}}$ (s$^{-1}$) & \hspace{4.5em}4.1(1)$\times$10$^{6}$\\
Natural linewidth $\Gamma/2\pi$ (MHz)      &  \hspace{4.5em}4.6(1)\\
Recoil velocity $\hbar k/ m$ (cm~s$^{-1}$) &  \hspace{4.5em}0.22\\
Doppler limit $\hbar \Gamma/2 k_B$ ($\mu$K) & \hspace{4.5em} 109(3)\\
Scattering acceleration $a_\gamma$ (km~s$^{-2}$) &  \hspace{4.5em}9.0(3)\\
\hline\hline
\end{tabular}
\end{table}

\begin{figure}
    \centering
    \includegraphics[width=0.5\textwidth]{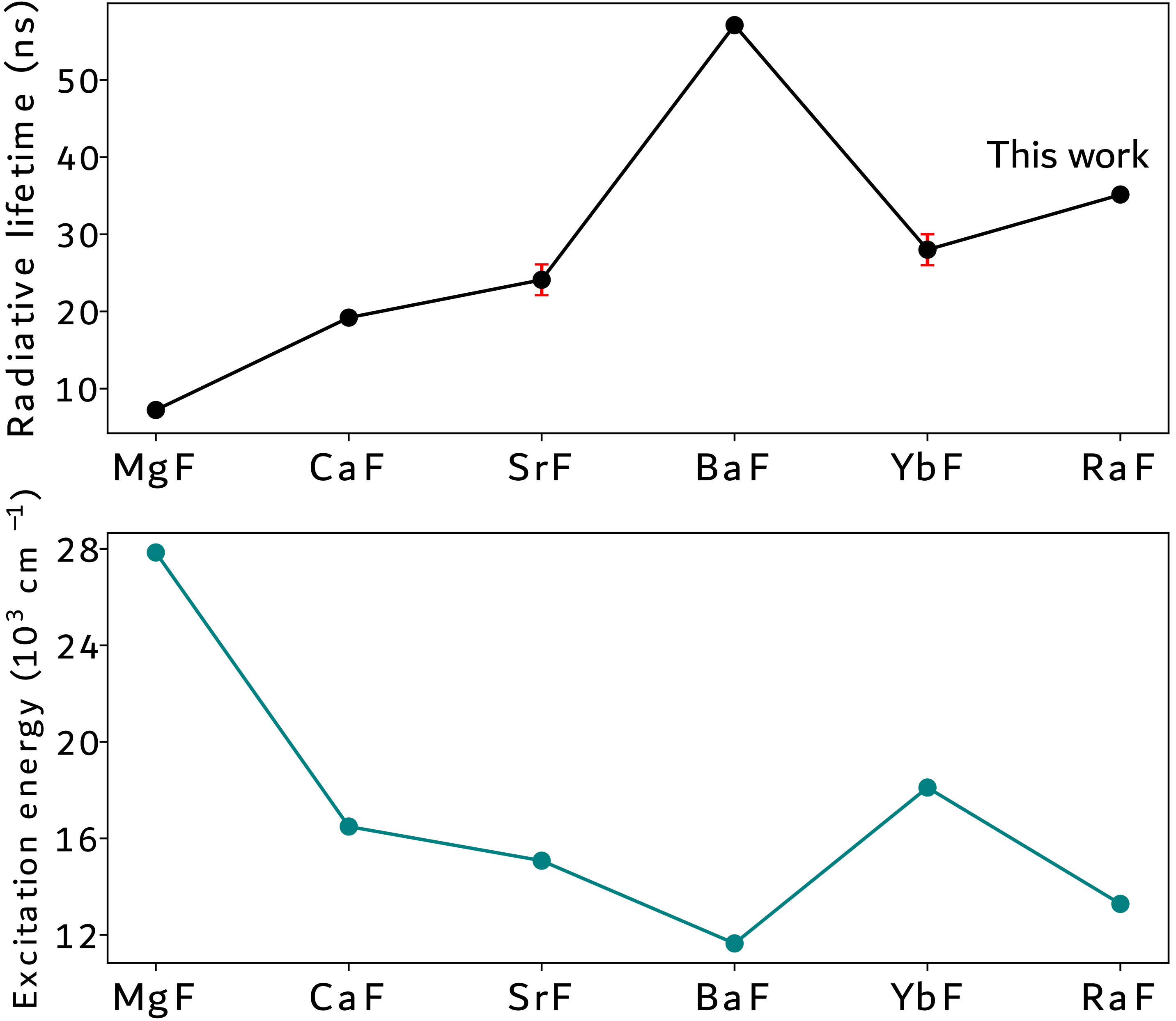}
    \caption{Comparison of the radiative lifetime (upper) and excitation energy (lower) of the $A$~$^2 \Pi_{1/2}$ state in RaF with MgF~\cite{Doppelbauer2022}, CaF~\cite{Wall2008,Kaledin1999}, SrF~\cite{Dagdigian1974,Hao2019}, BaF~\cite{Aggarwal2019,Barrow1988}, and YbF~\cite{Zhuang2011,Persinger2022}. In most cases, the error bar is smaller than the data marker.}
    \label{fig:comparison}
\end{figure}

Based on the lifetime $\tau = 35(1)$~ns, the radiative decay rate $\Gamma = \frac{1}{\tau} = 2.86(8)\times10^{7}$~s$^{-1}$ can be extracted. $\Gamma$ is a key quantity for a number of spectroscopic properties, such as the natural linewidth ($\Gamma/2\pi$) of the $A$~$^2 \Pi_{1/2} \leftarrow X$~$^2\Sigma_{1/2}$ transition, which is found to be equal to 4.6(1)~MHz, and the maximum photon scattering rate $R^{\rm{max}}_{\rm{sc}}$ of the transition.

For the laser-cooling scheme proposed in Ref.~\cite{Udrescu2023}, $R^{\rm{max}}_{\rm{sc}}$ can be approximated for {\color{black}optimistic} conditions of a magneto-optical trap as $R^{\rm{max}}_{\rm{sc}}=\Gamma/7$, {\color{black}based on a set of rate equations} as summarized in Ref.~\cite{Fitch2021LaserCooledMolecules}, which in this case is equal to $R^{\rm{max}}_{\rm{sc}}=4.1(1)\times10^6$~s$^{-1}$. With this scattering rate, the optically closed $A$~$^2 \Pi_{1/2} \leftarrow X$~$^2\Sigma_{1/2}$ transition in RaF could scatter 100,000 photons in 24.4(7)~ms, assuming identical $\Gamma$ for the $v=0,1,2$ states of $A$~$^2 \Pi_{1/2}$~\cite{Udrescu2023}. {\color{black}It is noted that the accuracy of this expression for $R^{\rm{max}}_{\rm{sc}}$ is limited by the validity of the rate model in a practical setting~\cite{Fitch2021LaserCooledMolecules}, and these results thus represent optimistic values. It is possible that experimental measurements of the scattering rate could be lower than this value.}

{\color{black}Lastly,} considering the photon recoil velocity of 0.22~cm~s$^{-1}$, the maximum scattering rate implies a photon scattering acceleration of $a_\gamma = R^{\rm{max}}_{\rm{sc}} \hbar k/m=9.0(3)$~km~s$^{-2}$. These results are summarized in Table~\ref{tab:summary}.

\begin{figure}
    \centering
    \includegraphics[width=0.45\textwidth]{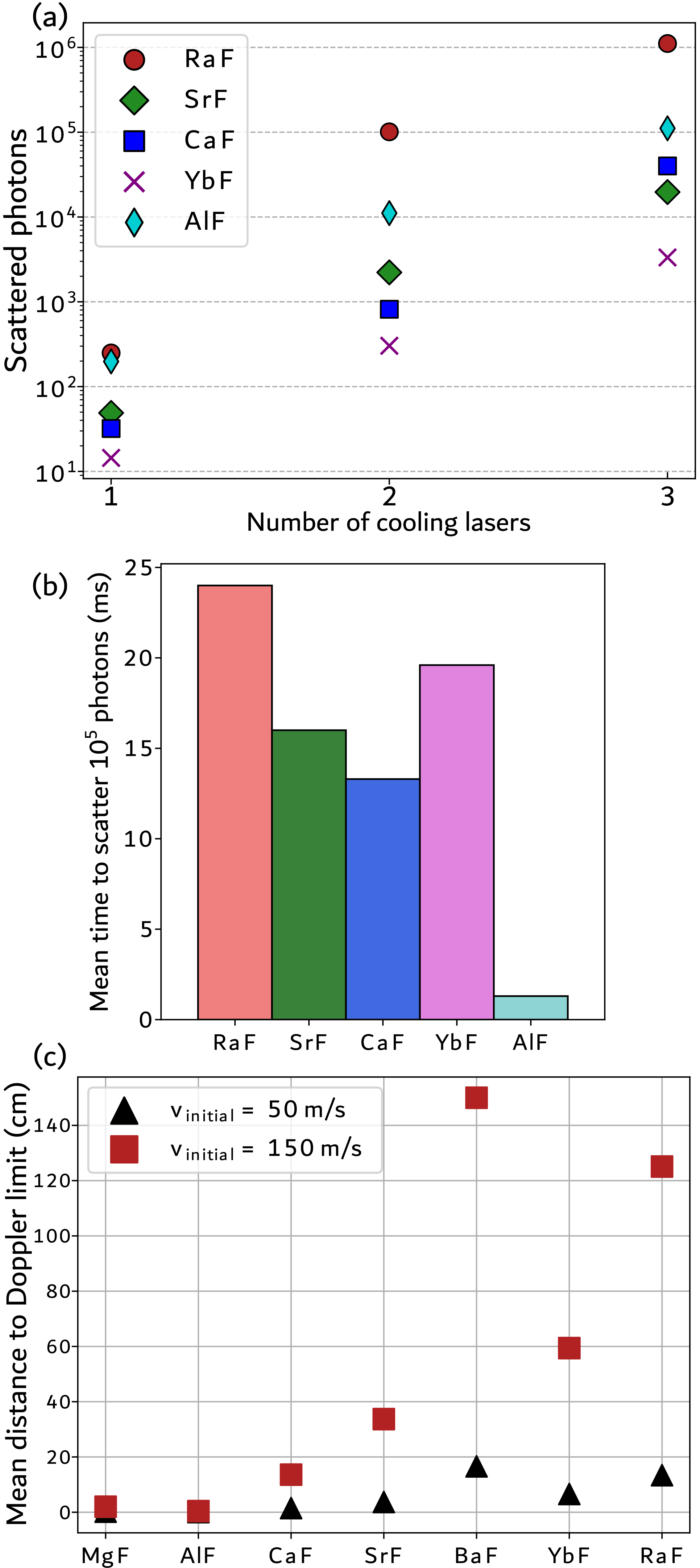}
    \caption{\textbf{(a)} Comparison of number of photons that can be scattered using the $A$~$\Pi$-$X$~$\Sigma$ transition before RaF, SrF, CaF, YbF, and AlF molecules would land in a dark state as a function of the number of cooling lasers. It is noted that laser-cooling in AlF is based on the $A$~$^1 \Pi$-$X$~$^1\Sigma$ transition, contrary to the $A$~$^2\Pi$-$X$~$^2\Sigma$ transition for the other molecules. \textbf{(b)} Average time to scatter 10$^5$ photons for the same molecules using the $A$~$\Pi$-$X$~$\Sigma$ laser-cooling cycle. Values were extracted using reported {\color{black}Franck-Condon factors} and radiative decay rates~\cite{Udrescu2023,Shuman2010,Truppe2017,Lim2018,Hofsass2021}, and the current result for the lifetime of $A$~$^2\Pi_{1/2}$ in RaF. \textbf{(c)} Comparison of mean distance required to slow to the Doppler limit a molecular beam produced by a cryogenic buffer-gas cell.}
    \label{fig:FigPhotonScattering}
\end{figure}

Fig.~\ref{fig:FigPhotonScattering} compares, {\color{black}based on the lifetime of the $A$~$^2\Pi_{1/2}$ state and the Franck-Condon factors}, the expected performance of laser cooling for RaF and stable diatomic molecules that have been successfully cooled, all featuring a laser-cooling cycle based on optically closed $\Sigma$-$\Pi$ transitions. Due to its highly diagonal Franck-Condon matrix for the $A$~$\Pi$-$X$~$\Sigma$ transition~\cite{Udrescu2023}, which far exceeds the diagonality of the transition in the other molecules, the $A$~$^2\Pi_{1/2}$-$X$~$^2\Sigma_{1/2}$ transition in RaF can scatter an order of magnitude more photons when using 3 cooling lasers before a significant fraction of the molecules is lost to higher vibrational states, while potentially scattering up to 10$^5$ photons using only 2 lasers that address the $v=0,1$ vibrational states, as shown in Fig.~\ref{fig:FigPhotonScattering}a.

Although the radiative decay rate $\Gamma$ is the smallest for RaF, the time required to scatter 10$^5$ photons, which can also serve as a measure of the time required to slow a molecular beam down to the capture velocity of a magneto-optical trap, is comparable to that for other molecules (Fig.~\ref{fig:FigPhotonScattering}b). Molecular beams produced by cryogenic buffer-gas cells, which most often serve as the source in molecular trapping setups, are emitted with forward velocities that appear to be weakly dependent on the molecular mass of the beam~\cite{Wright2023BufferGas,Hutzler2011}. Fig.~\ref{fig:FigPhotonScattering}c compares the distance required to slow a molecular beam down to the Doppler limit using counter-propagating laser cooling. For a forward velocity of 50~m/s, which is achievable with a two-stage buffer-gas cell at 1~K, the difference in slowing RaF compared to the lighter molecules becomes insignificant for a beamline of realistic dimensions.

As a result, the diagonality of the transition in RaF, which could allow effective laser cooling with fewer lasers, is accompanied by a scattering rate that is comparable to that of other laser-cooled molecules. 
A more conclusive discussion on the number of photons scattered by RaF {\color{black}per cooling laser} would require {\color{black}direct} measurements of vibrational branching ratios {\color{black}in the future}.

Beyond RaF, other radium-containing molecules have also been proposed for searches of new physics based on calculations of a laser-coolable electronic structure. 
{\color{black}Neutral polyatomic} radium-containing molecules 
appear to have lifetimes in the ns range, such as $\tau_{\rm{calc}}=40$~ns for RaOH~\cite{Isaev2017},  making them similarly promising candidates as RaF for laser cooling and trapping. Therefore, experimental methods for the measurement of symmetry-violating moments based on laser-slowed and -cooled neutral molecules~\cite{Fitch2021} are expected to be critical for progress in designing sensitive searches for new physics with radium-containing molecules.

\section{Conclusions}
In this work, the measurement of the radiative lifetime of the $A$~$^2 \Pi_{1/2}$ ($v=0$) state in RaF is presented. The lifetime was measured via delayed multi-step ionization using the CRIS experiment at the CERN-ISOLDE radioactive ion beam facility. The lifetime is determined to be $\tau = 35(1)$~ns. Using this value, the radiative decay rate $\Gamma =  2.86(8)\times10^{7}$~s$^{-1}$ is extracted, which further determines 
the maximum photon scattering rate $R^{\rm{max}}_{\rm{sc}} = 4.1(1)\times10^{6}$~s$^{-1}$. The validity of the lifetime measurement in RaF is benchmarked via a measurement of the lifetime of the $8P_{3/2}$ state in Fr using the same technique. The extracted lifetime for the $8P_{3/2}$ state in Fr at 83(3)~ns in this work is in agreement with the literature value.
%

The quantities reported in this work are of direct relevance to simulations for efficient laser cooling of RaF. Laser-cooled and trapped neutral molecules are envisioned as sensitives probes for future searches of new physics, and thanks to its exceptionally diagonal Franck-Condon matrix, RaF is expected to scatter 10$^5$ photons using only 2 cooling lasers, while using 3 cooling lasers allows scattering an order of magnitude more photons than other laser-cooled molecules. While the photon-scattering acceleration is lower in RaF compared to the lighter homoelectronic molecules, it is shown that the difference becomes insignificant for a molecular source that produces a sufficiently slow beam ($\sim$50~m/s).
\\
\section*{Acknowledgments}
We thank the ISOLDE technical teams for their support.

This project has received funding from the European Union's Horizon Europe Research and Innovation programme EUROLABS under Grant Agreement No. 101057511 and the European Union's Horizon 2020 research and innovation programme under Grant Agreement No.~654002.

Financial support from FWO, as well as from the Excellence of Science (EOS) programme (No. 40007501) and the KU Leuven project C14/22/104, is acknowledged. The STFC consolidated grants ST/V001116/1 and ST/P004423/1 and the FNPMLS ERC grant agreement No.~648381 are acknowledged. SGW and RFGR acknowledge funding by the Office of Nuclear Physics, U.S. Department of Energy Grants DE-SC0021176 and DE-SC002117. MAu, AR, JWa, and JWe acknowledge funding from the EU’s H2020-MSCA-ITN Grant No.~861198 ‘LISA’. DH acknowledges financial support from the Swedish Research Council~(2020-03505). JL acknowledges financial support from STFC grant ST/V00428X/1. SWB, YCL, WCM, and XFY acknowledge support from the National Natural Science Foundation of China (No.~12350007).



%

\end{document}



\title{Supplemental material: Radiative lifetime of the $A$~$^2 \Pi_{1/2}$ state in RaF with relevance to laser cooling}

\author{M.~Athanasakis-Kaklamanakis\orcidlink{0000-0003-0336-5980}}
 \email{m.athkak@cern.ch}
\affiliation{Experimental Physics Department, CERN, CH-1211 Geneva 23, Switzerland}
\affiliation{KU Leuven, Instituut voor Kern- en Stralingsfysica, B-3001 Leuven, Belgium}
\affiliation{Centre for Cold Matter, Imperial College London, SW7 2AZ London, United Kingdom}

\author{S.~G.~Wilkins}
\affiliation{Department of Physics, Massachusetts Institute of Technology, Cambridge, MA 02139, USA}
\affiliation{Laboratory for Nuclear Science, Massachusetts Institute of Technology, Cambridge, MA 02139, USA}

\author{{P.~Lass\`egues}}
\affiliation{KU Leuven, Instituut voor Kern- en Stralingsfysica, B-3001 Leuven, Belgium}

\author{L.~Lalanne}
\affiliation{Experimental Physics Department, CERN, CH-1211 Geneva 23, Switzerland}

\author{J.~R.~Reilly}
\affiliation{Department of Physics and Astronomy, The University of Manchester, Manchester M13 9PL, United Kingdom}

\author{{O.~Ahmad}}
\affiliation{KU Leuven, Instituut voor Kern- en Stralingsfysica, B-3001 Leuven, Belgium}

\author{M.~Au\orcidlink{0000-0002-8358-7235}}
\affiliation{Systems Department, CERN, CH-1211 Geneva 23, Switzerland}
\affiliation{Department of Chemistry, Johannes Gutenberg-Universit\"{a}t Mainz, 55099 Mainz, Germany}

\author{{S.~W.~Bai}}
\affiliation{School of Physics and State Key Laboratory of Nuclear Physics and Technology, Peking University, Beijing 100971, China}

\author{{J.~Berbalk}}
\affiliation{KU Leuven, Instituut voor Kern- en Stralingsfysica, B-3001 Leuven, Belgium}

\author{{C.~Bernerd}}
\affiliation{Systems Department, CERN, CH-1211 Geneva 23, Switzerland}

\author{A.~Borschevsky\orcidlink{0000-0002-6558-1921}}
\affiliation{Van Swinderen Institute of Particle Physics and Gravity, University of Groningen, Groningen 9712 CP, Netherlands}

\author{A.~A.~Breier\orcidlink{0000-0003-1086-9095}}
\affiliation{Institut f\"ur Optik und Atomare Physik, Technische Universit\"at Berlin, 10623 Berlin, Germany}
\affiliation{Laboratory for Astrophysics, Institute of Physics, University of Kassel, Kassel 34132, Germany}

\author{K.~Chrysalidis}
\affiliation{Systems Department, CERN, CH-1211 Geneva 23, Switzerland}

\author{T.~E.~Cocolios\orcidlink{0000-0002-0456-7878}}
\affiliation{KU Leuven, Instituut voor Kern- en Stralingsfysica, B-3001 Leuven, Belgium}

\author{R.~P.~de~Groote\orcidlink{0000-0003-4942-1220}}
\affiliation{KU Leuven, Instituut voor Kern- en Stralingsfysica, B-3001 Leuven, Belgium}

\author{C.~M.~Fajardo-Zambrano\orcidlink{0000-0002-6088-6726}}
\affiliation{KU Leuven, Instituut voor Kern- en Stralingsfysica, B-3001 Leuven, Belgium}

\author{K.~T.~Flanagan\orcidlink{0000-0003-0847-2662}}
\affiliation{Department of Physics and Astronomy, The University of Manchester, Manchester M13 9PL, United Kingdom}
\affiliation{Photon Science Institute, The University of Manchester, Manchester M13 9PY, United Kingdom}

\author{S.~Franchoo}
\affiliation{Laboratoire Ir\`{e}ne Joliot-Curie, Orsay F-91405, France}
\affiliation{University Paris-Saclay, Orsay F-91405, France}

\author{R.~F.~Garcia~Ruiz}
\affiliation{Department of Physics, Massachusetts Institute of Technology, Cambridge, MA 02139, USA}
\affiliation{Laboratory for Nuclear Science, Massachusetts Institute of Technology, Cambridge, MA 02139, USA}

\author{D.~Hanstorp\orcidlink{0000-0001-6490-6897}}
\affiliation{Department of Physics, University of Gothenburg, Gothenburg SE-41296, Sweden}

\author{R.~Heinke}
\affiliation{Systems Department, CERN, CH-1211 Geneva 23, Switzerland}

\author{{P.~Imgram}\orcidlink{0000-0002-3559-7092}}
\affiliation{KU Leuven, Instituut voor Kern- en Stralingsfysica, B-3001 Leuven, Belgium}

\author{\'{A}.~Koszor\'{u}s\orcidlink{0000-0001-7959-8786}}
\affiliation{Experimental Physics Department, CERN, CH-1211 Geneva 23, Switzerland}
\affiliation{KU Leuven, Instituut voor Kern- en Stralingsfysica, B-3001 Leuven, Belgium}

\author{A.~A.~Kyuberis\orcidlink{0000-0001-7544-3576}}
\affiliation{Van Swinderen Institute of Particle Physics and Gravity, University of Groningen, Groningen 9712 CP, Netherlands}

\author{{J.~Lim}\orcidlink{0000-0002-1803-4642}}
\affiliation{Centre for Cold Matter, Imperial College London, SW7 2AZ London, United Kingdom}

\author{{Y.~C.~Liu}}
\affiliation{School of Physics and State Key Laboratory of Nuclear Physics and Technology, Peking University, Beijing 100971, China}

\author{{K.~M.~Lynch}\orcidlink{0000-0001-8591-2700}}
\affiliation{Department of Physics and Astronomy, The University of Manchester, Manchester M13 9PL, United Kingdom}

\author{{A.~McGlone\orcidlink{0000-0003-4424-865X}}}
\affiliation{Department of Physics and Astronomy, The University of Manchester, Manchester M13 9PL, United Kingdom}

\author{{W.~C.~Mei}}
\affiliation{School of Physics and State Key Laboratory of Nuclear Physics and Technology, Peking University, Beijing 100971, China}

\author{G.~Neyens\orcidlink{0000-0001-8613-1455}}
\email{gerda.neyens@kuleuven.be}
\affiliation{KU Leuven, Instituut voor Kern- en Stralingsfysica, B-3001 Leuven, Belgium}

\author{{L.~Nies}\orcidlink{0000-0003-2448-3775}}
\affiliation{Experimental Physics Department, CERN, CH-1211 Geneva 23, Switzerland}

\author{A.~V.~Oleynichenko\orcidlink{0000-0002-8722-0705}}
\affiliation{Affiliated with an institute covered by a cooperation agreement with CERN.}

\author{{A.~Raggio}\orcidlink{0000-0002-5365-1494}}
\affiliation{Department of Physics, University of Jyv\"{a}skyl\"{a}, Jyv\"{a}skyl\"{a} FI-40014, Finland}

\author{S.~Rothe}
\affiliation{Systems Department, CERN, CH-1211 Geneva 23, Switzerland}

\author{L.~V.~Skripnikov\orcidlink{0000-0002-2062-684X}}
\affiliation{Affiliated with an institute covered by a cooperation agreement with CERN.}

\author{{E.~Smets}}
\affiliation{KU Leuven, Instituut voor Kern- en Stralingsfysica, B-3001 Leuven, Belgium}

\author{B.~van~den~Borne\orcidlink{0000-0003-3348-7276}}
\affiliation{KU Leuven, Instituut voor Kern- en Stralingsfysica, B-3001 Leuven, Belgium}

\author{{J.~Warbinek}}
\affiliation{GSI Helmholtzzentrum f\"ur Schwerionenforschung GmbH, 64291 Darmstadt, Germany}
\affiliation{Department of Chemistry - TRIGA Site, Johannes Gutenberg-Universit\"at Mainz, 55128 Mainz, Germany}

\author{J.~Wessolek\orcidlink{0000-0001-9804-5538}}
\affiliation{Department of Physics and Astronomy, The University of Manchester, Manchester M13 9PL, United Kingdom}
\affiliation{Systems Department, CERN, CH-1211 Geneva 23, Switzerland}

\author{X.~F.~Yang\orcidlink{0000-0002-1633-4000}}
\affiliation{School of Physics and State Key Laboratory of Nuclear Physics and Technology, Peking University, Beijing 100971, China}

\author{the ISOLDE Collaboration}

\date{\today}


\maketitle

\section{Extended methods}
The measurements in this work were obtained using the Collinear Resonance Ionization Spectroscopy (CRIS) experiment at the ISOLDE radioactive ion beam facility at CERN~\cite{Cocolios2013CRIS}. $^{226}$Ra ($T_{1/2}=1600$~years) and $^{221}$Fr ($T_{1/2}=4.9$~minutes) nuclides were produced by impinging 1.4-GeV protons onto a room-temperature uranium carbide target. {Two weeks after the end of proton irradiation, $^{226}$Ra$^{19}$F molecules were formed by heating the target up to 1700~$^\circ$C and introducing a controlled leak of CF$_4$ in the target container (more details on the formation and production procedure can be found in Ref.~\cite{Au2023InsourceIntrap}). While the half-life of $^{221}$Fr is shorter than the delay from irradiation to extraction, $^{221}$Fr is continuously produced within the target from the radioactive decay of long-lived parent nuclides.}

The neutral $^{226}$Ra$^{19}$F molecules and $^{221}$Fr atoms were ionized using a tungsten surface ion source heated to 2000~$^\circ$C, and the radioactive ion beam containing $^{226}$Ra$^{19}$F$^+$ and $^{221}$Fr$^+$ among all other radioactive species was accelerated to 40,000~eV. 

\begin{figure}
    \centering
     \includegraphics[width=0.45\textwidth]{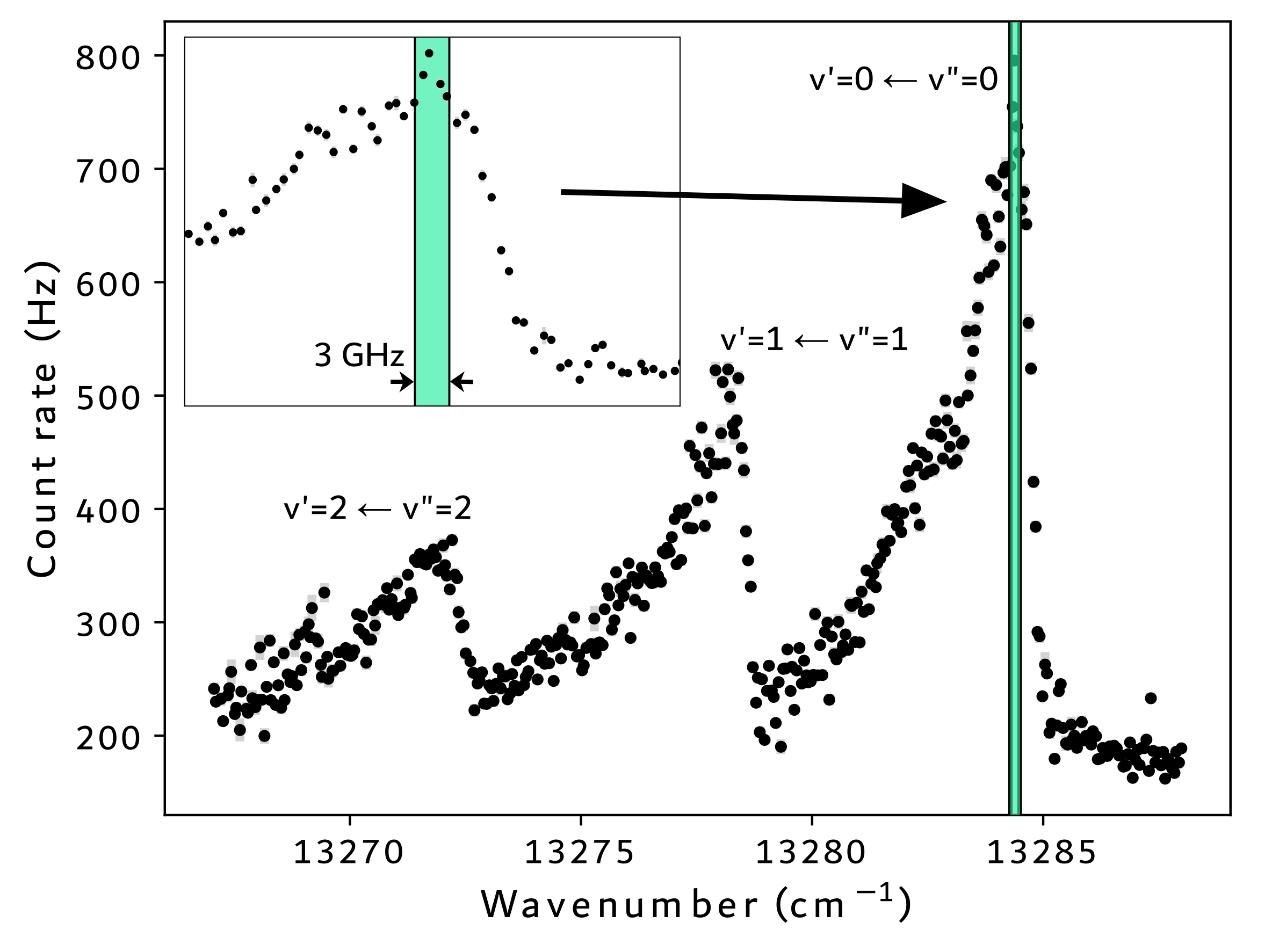}
    \caption{Spectrum of the $A$~$^2\Pi_{1/2} \leftarrow X$~$^2\Sigma_{1/2}$ transition measured with the broadband laser used for the excitation step in this study (753-nm in Fig.~1 of main text). The excitation laser was fixed at the center of the wavenumber range denoted by the shaded region, whose width corresponds to the laser's linewidth.}
    \label{fig:FigSpectrum}
\end{figure}

The beam of $^{226}$Ra$^{19}$F$^+$ molecular ions was separated from all other radioactive ions using two successive dipolar magnetic separators. When a beam of $^{221}$Fr$^+$ was required, the current on the magnetic separators was appropriately tuned to allow transmission of species with a mass-to-charge ratio $A/q=221$ instead of $A/q=245$. The isotopically pure beam of $^{226}$Ra$^{19}$F$^+$ or $^{221}$Fr$^+$ was then injected into a gas-filled linear Paul trap~\cite{Mane2009} where the continuous beam was accumulated, cooled to room temperature by colliding with He buffer gas, and released in bunches of 5-$\mu$s temporal width with a repetition rate of 100~Hz. The bunches were released with a kinetic energy that drifted over time between 39,905 and 39,910~eV and the acceleration voltage was monitored with a 7.5-digit digital multimeter (Keithley DMM7510) with a precision of 100 mV.


\begin{figure*}
    \centering
    \includegraphics[width=0.85\textwidth]{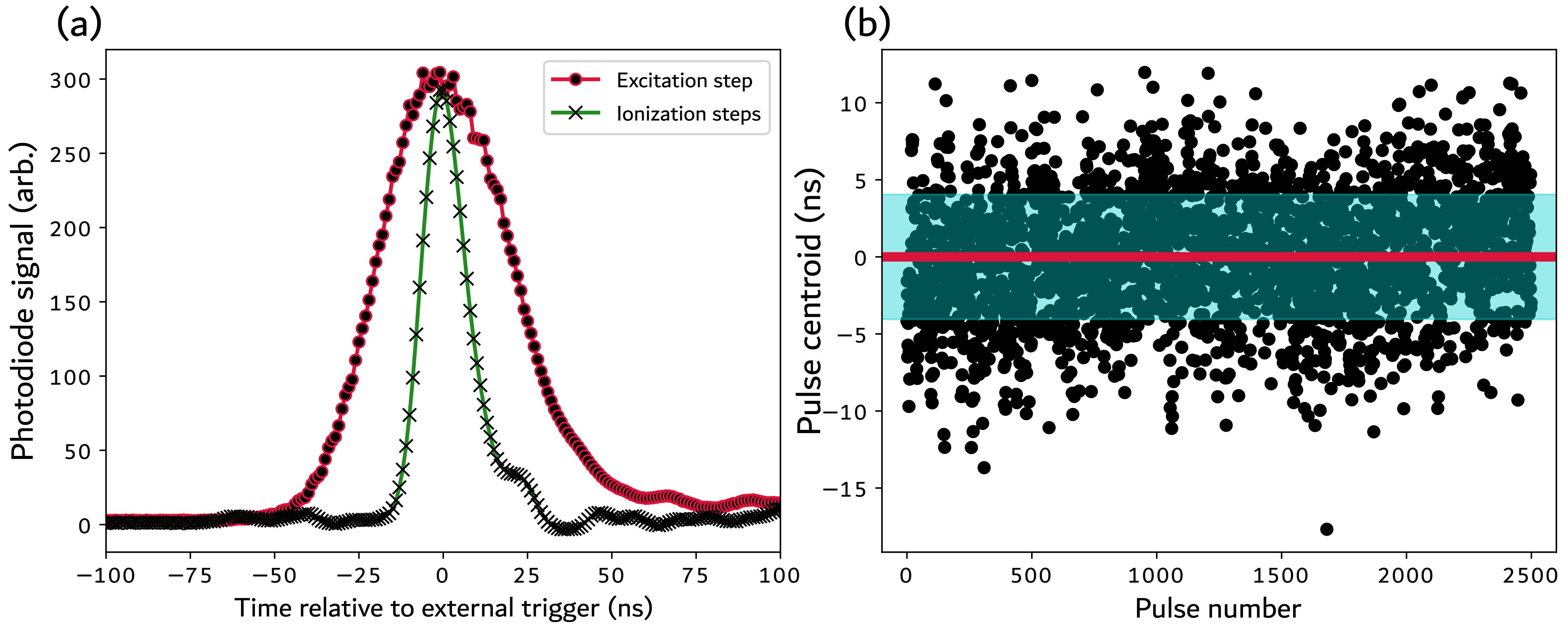}
    \caption{\textbf{(a)} Measured time profiles of the pulses of the lasers used for the excitation and ionization steps in the setup, averaged across 20 pulses. \textbf{(b)} Centroid with respect to an external, ultra-low-jitter (50~ps) trigger for 2,500 pulses of the Ti:Sa laser. The horizontal line and surrounding band represent the mean and standard deviation of the distribution, respectively. The standard deviation is taken as the pulse-to-pulse jitter of the laser, and is equal to 4~ns for the Ti:Sa laser.}
    \label{fig:FigPulseProfiles}
\end{figure*}

For RaF, the wavelength of one laser beam (titanium sapphire, 1-kHz repetition rate, 3-GHz linewidth) was centered at the \textit{Q}(13) line of the $A$~$^2 \Pi_{1/2} \leftarrow X$~$^2 \Sigma_{1/2}$ ($v'=0 \leftarrow v''=0$) (753~nm) transition from the ground state, which corresponds to the most intense part of the broadband spectrum, see Fig.~\ref{fig:FigSpectrum}. Due to the broad linewidth of the excitation laser ($\sim$3~GHz), tens of overlapped rotational lines in the vicinity of $Q$(13) were being excited as well. A second laser beam (pulsed dye laser, 100-Hz repetition rate, $\sim$9-GHz linewidth, pumped by 532-nm Nd:YAG) was tuned to the frequency of the $G$~$^2 \Pi_{1/2} \leftarrow A$~$^2 \Pi_{1/2}$ ($v'=0 \leftarrow v''=0$) (646~nm) transition~\cite{AthKak2023Excited} as a subsequent excitation step. The third laser beam (532-nm Nd:YAG, 100-Hz repetition rate) non-resonantly ionized the molecules that had been excited to the $G$~$^2 \Pi_{1/2}$ $v=0$ state. The resulting ions were steered away from the residual neutral molecules and onto a single-ion detector. For the benchmark measurements with $^{221}$Fr, a two-step scheme was employed (see Fig.~1 of main text), using the second harmonic of the Ti:Sa laser used for RaF for the transition from the $7S_{1/2}$ ground state to the $8P_{3/2}$ excited state (423~nm) and the fundamental wavelength of the Nd:YAG laser (1064~nm) for the non-resonant ionization step.

\begin{figure*}
    \centering
    \includegraphics[width=0.85\textwidth]{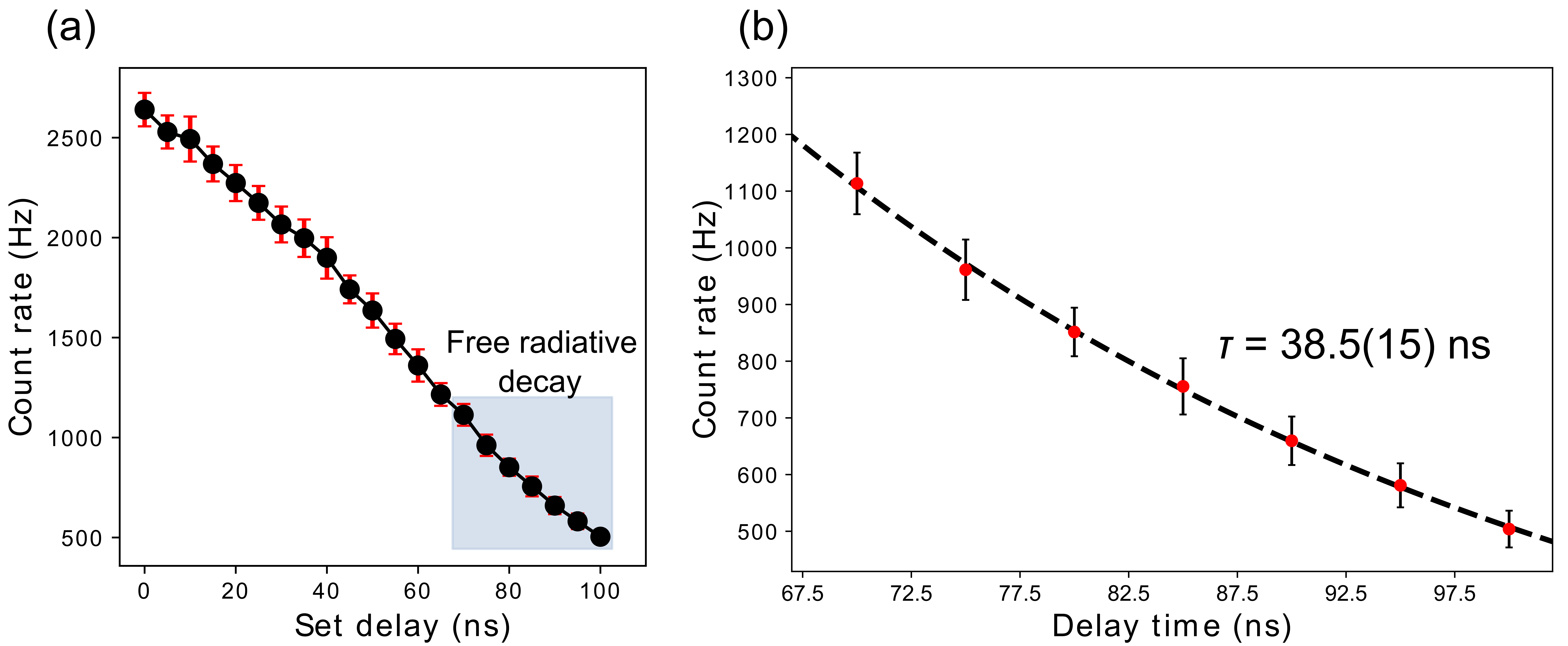}
    \caption{\textbf{(a)} Decay curve measured during the proof-of-principle measurement cycle. The shaded region denotes the part of the curve that was determined to correspond to the free radiative decay regime where the excitation and ionization lasers are not overlapped. \textbf{(b)} Fit of the exponential decay function to the experimental measurements. {The \textit{x}-error (not shown, included in Fig.~\ref{fig:SM_curves_res}) is common for all points and corresponds to the relative pulse-to-pulse laser jitter, equal to 5~ns.}}
    \label{fig:previous_campaign}
\end{figure*}

The pulse timing profiles of the lasers and their pulse-to-pulse timing stability (jitter) were measured prior to the experiment using a PicoScope 3000 oscilloscope and the signals of three Si photodiode detectors placed on a laser table before the start of the beamline. The pulse profiles were determined by averaging the signal of each photodiode (one photodiode receiving light from only one laser, also ensuring the photodiodes are not saturated) across 20 pulses, as registered by the externally triggered oscilloscope. The timing curves (see Fig.~\ref{fig:FigPulseProfiles}) were then fitted with skewed Voigt profiles to determine their full-width at half-maximum (FWHM). By collecting the timing profiles (individual pulses, not averaged) of 2,500 pulses for each laser, fitting them, and extracting the centroids, the jitter of each laser was also measured as the standard deviation of the pulse timing distribution.

\begin{table}[]
\caption{Summary of the details of the different cycles in the lifetime measurement of the $8P_{3/2}$ state in Fr. The systematic $y$-error corresponds to the added uncertainty on the count rate to account for atomic beam fluctuations.}
\label{tab:Fr_supp}
\begin{tabular}{ccccc}
 \hline \hline
Cycle &  \begin{tabular}[c]{@{}c@{}}Step\\timing\\ modulated\end{tabular} &\begin{tabular}[c]{@{}c@{}}Paul-trap\\ ejection\\ detuning ($\mu$s)\end{tabular} & \begin{tabular}[c]{@{}c@{}}Syst.\\ y-error\\ ($\%$)\end{tabular}  & \begin{tabular}[c]{@{}c@{}}$8P_{3/2}$\\ lifetime\\ (ns)\end{tabular} \\ \hline
1  & Ionization  & 0     & 3.2  & 81(6)              \\
2  & Ionization  & -0.5  & 3.1  & 78(7)              \\
3  & Ionization  & 0.5   & 1.1  & 101(7)             \\
4  & Excitation  & 0     & 5.9  & 78(5)              \\
   &             &       & \multicolumn{1}{r}{weighted mean:} & \textbf{83(3)}  \\
    \hline \hline
\end{tabular}
\end{table}

\begin{table}[]
\caption{Summary of the details of the different cycles in the lifetime measurement of the $A$~$^2\Pi_{1/2}$~$v=0$ state in RaF. The systematic $y$-error corresponds to the added uncertainty on the count rate to account for molecular beam fluctuations.}
\label{tab:RaF_supp}
\begin{tabular}{ccccc}
\hline\hline
Cycle                & \begin{tabular}[c]{@{}c@{}}Step\\timing\\modulated\end{tabular} & \begin{tabular}[c]{@{}c@{}}Paul-trap\\ ejection\\ detuning ($\mu$s)\end{tabular} & \begin{tabular}[c]{@{}c@{}}Syst.\\ y-error\\ ($\%$)\end{tabular} & \begin{tabular}[c]{@{}c@{}}$A$~$^2\Pi_{1/2}$\\ lifetime\\ (ns)\end{tabular} \\
\hline
1 & Ionization  & 0  & 1.5 & 34.1(1.2)\\
2 & Ionization  & -1 & 5.0 & 35.2(1.9)\\
3 & Ionization  & +1 & 0.9 & 32.9(1.4)\\
4 & Ionization  & 0  & 4.9 & 38.5(1.5)\\
5 & Excitation  & 0  & 4.6 & 34.5(2.6)\\
  &             &    & \multicolumn{1}{r}{weighted mean:} & \textbf{35(1)} \\
    \hline \hline                                                 
\end{tabular}
\end{table}

The pulses of the first step (excitation step) had a FWHM of 38(2)~ns, while the pulses of the second and third steps (ionization steps) had a FWHM of 15(1)~ns. The jitter of the excitation step was measured at 4~ns and that of the ionization steps at 3~ns, for a total relative jitter of 5~ns {(added in quadrature)} between excitation and ionization steps. As the second step in the RaF scheme was provided by a pulsed dye laser pumped by an identical 532-nm Nd:YAG laser as the one used for the non-resonant ionization step, the FWHM and jitter of the second and third steps were measured to be the same, as expected.

The laser pulse timings were controlled using an ultra-low-jitter ($<$50~ps) multi-channel signal generator (9520 Pulse Generator by Quantum Composers). During the measurements in RaF, the laser pulses of the two ionization steps (646-nm and 532-nm) were overlapped in time, and their timing with respect to the excitation step was controlled to introduce an ionization delay. Starting with the peaks of all three laser pulses being overlapped in time (determining the 0-ns delay point), the resonant ion count rate was measured as a function of the delay between the excitation and ionization steps. To measure the lifetime of the $A$~$^2\Pi_{1/2}$ state in RaF, the ionization steps were delayed in steps of 10~ns with respect to the excitation step. For the Fr benchmark measurements, the ionization step was delayed in steps of 20~ns from the excitation step.

For each lifetime measurement involved in this study, referred to as a measurement cycle, a total of 20 equidistant delay values were used. For each delay, all experimental parameters were kept fixed and the counts on the ion detector were recorded for 60 seconds. The counts recorded within the 60-s window were summed and taken as the total counts for the given delay, and the square root of the total counts was taken as the statistical error. The total counts and the corresponding uncertainty are thus expressed as a count rate per minute (units of min$^{-1}$).

\begin{figure*}
    \centering
\includegraphics[width=0.8\textwidth]{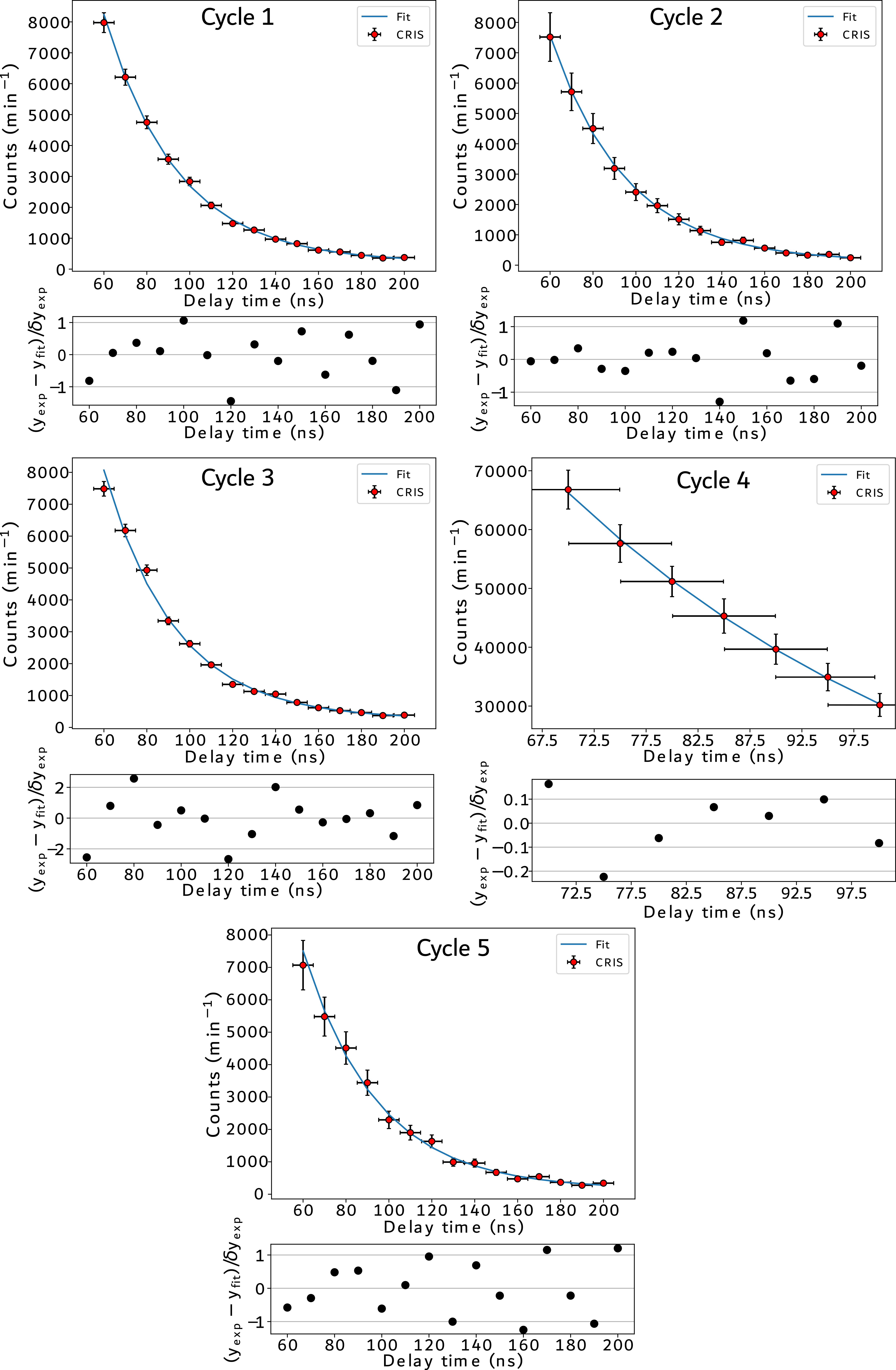}
    \caption{{Decay curves and fit residuals for the lifetime of the $A$~$^2\Pi_{1/2}$ state in RaF across all cycles.}}
    \label{fig:SM_curves_res}
\end{figure*}

In each measurement cycle, after taking measurements for 5 delay values, the timing was returned to the zero-delay and a reference measurement was taken to trace potential fluctuation and drifts in the molecular beam intensity over time. For each measurement cycle, a constant-parameter fit was made across all zero-delay reference \textit{y}-values and their errors. The percentage relative error in the fitted constant parameter for the reference measurements was then added to the statistical error of the counts for each delay value in the measurement cycle, representing a systematic scattering error due to the changing beam intensity.

Cycle 4 in Fig.~2c of the main text reports the lifetime value extracted from an experimental campaign that took place 2 years prior to the rest of the measurement cycles. This measurement was performed as a proof-of-principle run to confirm the suitability of the delayed resonance ionization technique for lifetime measurements.

In this proof-of-principle measurement, a total of 21 delays were set up to 100~ns, in steps of 5~ns. The startpoint analysis for the decay curve (Fig.~\ref{fig:previous_campaign}) determined that the free radiative decay regime begins between the 65- and 70-ns delay values. The middle point of 67.5~ns was thus chosen. Fitting the exponential decay function (Eq.~1 in main text) to the decay curve yielded a lifetime of $\tau = 38.5(15)$~ns.

{Fig.~\ref{fig:SM_curves_res} shows the decay curves and fit residuals for all the measurement cycles in the determination of the $A$~$^2\Pi_{1/2}$ $v=0$ lifetime in RaF. The lifetime values from the individual cycles and the overall weighted mean is given in Tables~\ref{tab:Fr_supp} and \ref{tab:RaF_supp}. For the constant fits, each point was weighted by the inverse of its uncertainty squared. The weighted mean was calculated with the formula:}
\begin{equation}
    \mu_w = \frac{\sum x_i/\sigma_i^2}{\sum 1/\sigma_i^2}
\end{equation}
{where $x_i$ is the nominal lifetime value from each individual cycle and $\sigma_i$ is the uncertainty. The error on the weighted mean is extracted as:}
\begin{equation}
    \sigma_w = \frac{1}{\sqrt{\sum 1/\sigma_i^2}}
\end{equation}

For the laser-cooling comparison across different molecules in Fig.~5 of the main text, the number of scattered photons as a function of number of cooling lasers was calculated using the reported branching ratios of the $v=0,1,2$ vibrational states of $A$~$^2\Pi_{1/2}$ in CaF~\cite{Truppe2017}, SrF~\cite{Shuman2010}, YbF~\cite{Lim2018}, and RaF~\cite{Udrescu2023}, and the $A$~$^1\Pi$ state in AlF~\cite{Hofsass2021}.

For 1 cooling laser, the number of scattered photons was calculated as the inverse of the sum of branching ratios to $v''=1,2,3$ of the lower state, starting from $v'=0$ of the upper state. For 2 cooling lasers, the sum of branching ratios to $v''=2,3$ starting from $v'=0,1$ were included, while for 3 cooling lasers, the sum of the branching ratios to $v''=3$ from $v'=0,1,2$ was included.


%